\documentclass[12pt]{iopart}
\usepackage{iopams}
\usepackage{graphicx}
\usepackage{pstricks}
\usepackage{subfigure}
\usepackage[english]{babel}
\usepackage{float}
\usepackage{color}
\usepackage[percent]{overpic}
\usepackage{cite} 
\usepackage{citesort} 

\newcommand\rnumber{\mathop{\mbox{$\rho$-$C$}}}

\begin{document}

\title[]{Interplay of disorder and $\mathcal{PT}$-symmetry}

\author{C. Mej\'{\i}a-Cort\'{e}s and M. I. Molina}

\address{Departamento de F\'{\i}sica, Facultad de Ciencias, Universidad de
Chile, Casilla 653, Santiago, Chile, Center for Optics and Photonics (CEFOP) and
MSI-Nucleus on Advanced Optics, Casilla 4016, Concepcion, Chile}
\ead{ccmejia@googlemail.com} 

\begin{abstract} 
We examine a one-dimensional $\mathcal{PT}$-symmetric binary
lattice in the presence of diagonal disorder. We focus on the wave transport
phenomena of localized and extended input beams for this  disordered system.
In the pure $\mathcal{PT}$-symmetric case, we derive an exact expression
for the evolution of light localization in terms of the typical parameters of
the system. In this case localization is enhanced as the gain and loss
parameter in increased. In the presence of disorder, we observe that the
presence of gain and loss inhibits (favors) the transport for localized
(extended) excitations.  
\end{abstract}

\vspace{2pc}

\maketitle

\section{Introduction} 

In 1958 Anderson showed, within the independent electron framework, that the
presence of a finite concentration of linear uncorrelated disorder completely
inhibits the quasiparticle propagation in one-dimension (1D) and two-dimension
(2D), giving rise to a saturation of its mean-square displacement and an
exponential decrease of the transmissivity of plane waves with system
size~\cite{PhysRev.109.1492,PhysRevLett.22.1065,doi:10.1080/14786436908216338}.
Proposed originally for electrons and one-particle excitations in solids~
\cite{PhysRev.109.1492,lif1988introduction,sheng1990scattering,2006iwsl.book.....P},
it was soon extended to many other fields such as
acustics~\cite{Weaver1990129,1.399867}, Bose-Einstein
condensates~\cite{billy2008direct} and
optics~\cite{PhysRevLett.53.2169,Freilikher1992137,wiersma1997localization,cite-key,PhysRevLett.100.013906,1367-2630-15-1-013045,nphand}.

A different and novel concept that has gained much recent attention is that of
$\mathcal{PT}$-symmetry. It is based on the seminal work of Bender and
coworkers~\cite{PhysRevLett.80.5243,PhysRevLett.89.270401}, who showed that
non-hermitian Hamiltonians are capable of displaying a purely real eigenvalue
spectrum, provided the system is invariant with respect to the combined
operations of parity ($\mathcal{P}$) and time-reversal ($\mathcal{T}$)
symmetry. For one-dimensional systems the $\mathcal{PT}$ requirement leads to
the condition that the imaginary part of the potential term in the Hamiltonian
be an odd function, while its real part be even. In a $\mathcal{PT}$-symmetric
system, the effects of loss and gain can balance each other and, as a result,
give rise to a bounded dynamics. The system thus described can experience a
spontaneous symmetry breaking from a $\mathcal{PT}$-symmetric phase (all
eigenvalues real) to a broken phase (at least two complex eigenvalues), as the
imaginary part of the potential is increased. In the case of optics, the
paraxial wave equation has the form of a Schr\"{o}dinger equation and, as a
consequence,  the potential is proportional to the index of refraction. The
$\mathcal{PT}$-symmetry requirements lead to the condition that the real part
of the refractive index be an even function, while the imaginary part be an odd
function in space. To date, numerous $\mathcal{PT}$-symmetric systems have been
explored in several fields, from
optics~\cite{El-Ganainy:07,PhysRevLett.100.030402,PhysRevLett.100.103904,PhysRevLett.103.093902,ruter2010observation,ptnat},
electronic circuits~\cite{PhysRevA.84.040101}, solid-state and atomic
physics~\cite{PhysRevLett.77.570,PhysRevA.82.030103}, to magnetic
metamaterials~\cite{PhysRevLett.110.053901}, among others.  The
$\mathcal{PT}$-symmetry-breaking phenomenon has been observed in several
experiments~\cite{ruter2010observation,PhysRevLett.103.093902,PhysRevA.84.021806,PhysRevA.84.012123}.

It is known that a 1D simple periodic lattice with homogeneous couplings and
endowed with gain and loss, displaying in this way $\mathcal{PT}$-symmetry, is
always in the broken phase of this symmetry and does not have a stable
parameter window~\cite{tsironis}. For finite $\mathcal{PT}$-symmetrical
lattices, it has been shown that $\mathcal{PT}$-symmetry is preserved inside a
parameter window whose size shrinks with the number of lattice
sites~\cite{PhysRevE.89.033201}. If one breaks the homogeneity of the
couplings, and consider an infinite binary lattice, it was shown that there is
a well-defined parameter window where $\mathcal{PT}$-symmetry is
preserved~\cite{Dmitriev:10}.

A previous study of the effect of $\mathcal{PT}$-symmetry on Anderson
localization, carried out on a (continuous) 2D square optical lattice, suggests
that the presence of $\mathcal{PT}$-symmetry enhances light
localization~\cite{Jovic:12}.  Recently, it has been observed that the presence
of $\mathcal{PT}$-symmetry in a (discrete) 1D waveguide array with binary
coupling give rise to light localization, i.e., ``emulate''
disorder~\cite{ptncom}.

In this work we are interested in examining the interplay between the
simultaneous presence of disorder and $\mathcal{PT}$-symmetry, and how this
affects the transport properties of extended excitations (plane waves) and the
dynamical evolution of a completely localized excitation across a 1D binary
lattice. We found that, for a disordered binary lattice,  the presence of gain
and loss tends to favor (inhibit) the transport of extended (localized)
excitations.

\section{The model}

\begin{figure}
\begin{center}
\begin{overpic}[width=0.85\textwidth]{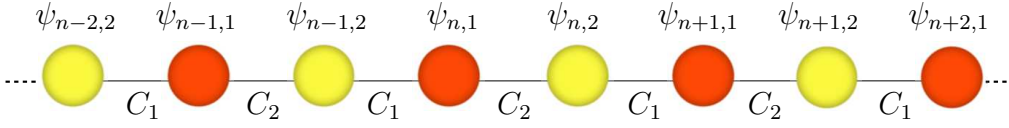}
\put(12,0){$C_1$}\put(3,9){$\scriptsize{\psi_{n-2,2}}$}
\put(24,0){$C_2$}\put(15,9){$\scriptsize{\psi_{n-1,1}}$}
\put(36,0){$C_1$}\put(28,9){$\scriptsize{\psi_{n-1,2}}$}
\put(49,0){$C_2$}\put(42,9){$\scriptsize{\psi_{n,1}}$}
\put(62,0){$C_1$}\put(54,9){$\scriptsize{\psi_{n,2}}$}
\put(74,0){$C_2$}\put(65,9){$\scriptsize{\psi_{n+1,1}}$}
\put(87,0){$C_1$}\put(77,9){$\scriptsize{\psi_{n+1,2}}$}
\put(90,9){$\scriptsize{\psi_{n+2,1}}$}
\end{overpic}
\caption{Sketch of the 1D linear binary lattice with alternating gain (yellow
filled circles) and loss (orange filled circles).}
\label{f1}
\end{center}
\end{figure}

Let us consider a weakly-coupled array of optical waveguides with binary
couplings (cf.~Fig.~\ref{f1}). In addition, each guide possesses a propagation
constant whose real part can be random, and whose imaginary part is distributed
across the array in a manner that satisfies the requirements of
$\mathcal{PT}$-symmetry, that is, the gain (yellow circles) or loss (orange 
circles)  coefficient alternates in sign from site to site. Such system can be
modeled by a set of coupled, discrete linear Schr\"odinger equations.
Considering only coupling between nearest-neighbor waveguides, the model is
described by:
\begin{equation}
\eqalign{
i\frac{\rmd\psi_{n,1}}{\rmd z}+C_1\psi_{n-1,2}+C_2\psi_{n,2}+(\gamma_{n,1} + i\rho_{n,1}) \psi_{n,1}&=0,\cr
i\frac{\rmd\psi_{n,2}}{\rmd z}+C_1\psi_{n+1,1}+C_2\psi_{n,1}+(\gamma_{n,2} - i\rho_{n,2})\psi_{n,2}&=0,}
\label{eq1}
\end{equation}
with $\gamma_{n,1(2)}=1+\varepsilon_{n,1(2)}$. Here $\varepsilon_{n,1(2)}$ is a
real random number and $\varepsilon_{n,1(2)}\in [-W/2, W/2]$ where $W$ is the
disorder width. A possible choice for the gain and loss coefficient
$\rho_{n,1(2)}$ is to set $\rho_{n,1}=+\rho$ and $\rho_{n,2}=-\rho$.

The optical power content for such a system is defined as
\begin{equation}
P=\sum_n{\left|\psi_{n,1}\right|^2+\left|\psi_{n,2}\right|^2}.
\label{eq2} 
\end{equation}
and in the absence of gain and loss, $P$ is a conserved quantity. Model~(\ref{eq1}) 
is a Hamiltonian system, where $i\rmd_z \psi_{n,1(2)}= \partial
H/\partial \psi_{n,1(2)}^{*}$. The (non-hermitian) Hamiltonian is given by
\begin{eqnarray}
H &=& \sum_{n} \left[ i \rho (|\psi_{n,1}|^2 - |\psi_{n,2}|^2) + 
C_{2} \psi_{n,1}^{*} \psi_{n,2} + \right.\nonumber\\
& & \left. C_{1} \psi_{n,1}^{*} \psi_{n-1,2} +  
C_{2} \psi_{n,2}^{*} \psi_{n,1}+ C_{1} \psi_{n,2}^{*} \psi_{n+1,1} \right].
\end{eqnarray}
%

In order to distinguish the spatial distribution (structure) of various
solutions, an useful quantity called the participation
rate of a solution $\psi_{n,1(2)}$ is defined as
\begin{equation}
R=\frac{P^2}{\sum_n{\left|\psi_{n,1}\right|^4+\left|\psi_{n,2}\right|^4}},
\label{eq3} 
\end{equation}
which indicates how many sites are effectively excited in the lattice. Here
$n$ runs over a half of the total number of sites ($N$). For a completely extended
state, $R=N$, while in the presence of complete localization, $R=1$.

We begin by looking at the structure of the modes of the corresponding
eigenvalue problem. As a first, and very rough preliminary view, we collapse
the whole lattice to only two sites, i.e., a dimer, and examine the behaviour
of the instability gain of the modes as a function of the gain and loss
parameter, and also as a function of the disorder width.

\section{The simplified dimer model}

The corresponding equations for the dimer model in our system are
\begin{equation}
i {\rmd \psi_{1}\over{\rmd z}} + (\varepsilon_{1}+ i \rho) \psi_{1} + C \psi_{2} = 0,\hspace{1cm}
i {\rmd \psi_{2}\over{\rmd z}} + (\varepsilon_{2}- i \rho) \psi_{2} + C \psi_{1} = 0.
\end{equation}
We look for stationary solutions $\psi_{1(2)}(z) \sim \psi_{1(2)}\exp{(i \lambda
z)}$. This leads to the eigenvalue equation
\begin{equation}
(-\lambda + \varepsilon_{1}+ i \rho) \psi_{1} + C \psi_{2}= 0,\hspace{1cm}
(-\lambda + \varepsilon_{2}- i \rho) \psi_{2} + C \psi_{1}= 0.
\label{eq:6}
\end{equation}
After solving the eigenvalue problem, one obtains the propagation constant
\begin{equation}
\lambda = {(\varepsilon_{1}+\varepsilon_{2})\over{2}} \pm {1\over{2}} \sqrt{(\varepsilon_{1}-
\varepsilon_{2})^2 - 4 \rho^2 + 4 C^2 + 4 i (\varepsilon_{1}-\varepsilon_{2})\rho}\label{eq:6b}
\end{equation}
In this oversimplified model, the disorder width is given by
$|\varepsilon_{1}-\varepsilon_{2}|$.

\begin{figure}
\begin{center}
\noindent
\begin{overpic}[width=1\textwidth]{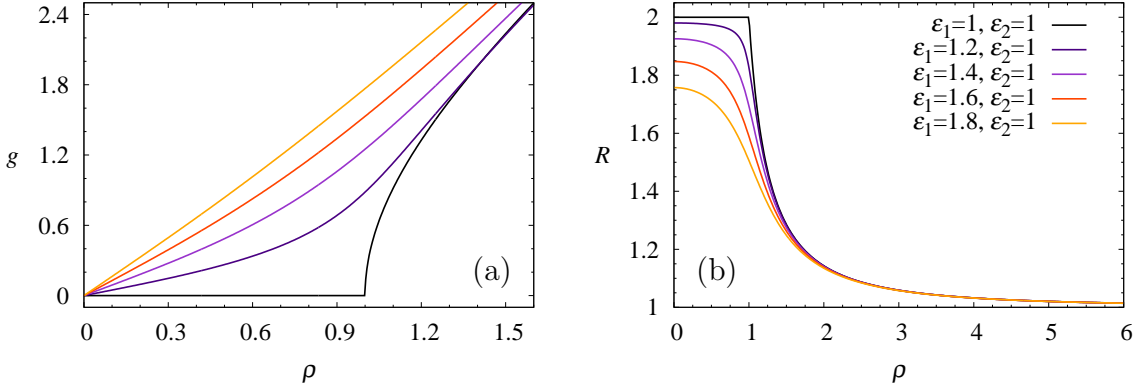}\put(41,9){(a)}\put(60,9){(b)}
\end{overpic}
\caption{(a) Instability gain $g$ and (b) participation ratio $R$ in the dimer model,
as a function of the gain and loss parameter $\rho$, both for several ``disorder
widths'' labeled at the inset. $C=1$.}
\label{f2}
\end{center}
\end{figure}

We note that $\lambda$ is in general a complex number, but in the absence of
``disorder'', i.e., when $\varepsilon_{1}=\varepsilon_{2}$, the system is
$\mathcal{PT}$-symmetric and there is a parameter window where $\lambda\in\Re$:
$\rho<C$. We conclude that the presence of any amount of disorder gives rise to
a complex propagation constant. Now, let us look at the behaviour of the
imaginary part of $\lambda$ as a function of $\rho$, keeping the coupling
constant, $C=1$. From Eq.~(\ref{eq:6b}) we obtain the imaginary part of
$\lambda$, or instability gain, as
\begin{equation}
g = \frac{1}{\sqrt{2}}\left(-a + \sqrt{a^2 + b^2}\right)^{1/2}
\end{equation}
where, $a=(\varepsilon_{1}-\varepsilon_{2})^2 - 4 \rho^2 + 4 C^2$, $b=4
(\varepsilon_{1}-\varepsilon_{2}) \rho$. Figure~\ref{f2}(a) shows the behaviour of $g$
as a function of $\rho$ for several values of disorder width, $W$. 

Perhaps the most interesting feature of this graph is the fact that the
instability gain increases as a function of disorder, for a fixed gain and loss
parameter. At large enough $\rho$ values, all the curves fall eventually on the
$W=0$ case: $g=\Theta(\rho-1) \sqrt{\rho^2-1}$, where $\Theta(x)$ is the step
function: $\Theta(x)=0$ for $x<0$, or $\Theta(x)=1$ for $x>0$. 

Now, let us look at the behaviour of the participation ratio $R$ for our dimer
system:
\begin{equation}
R = {(1 + \alpha^2)^2\over{1+ \alpha^4}}
\end{equation}
where $\alpha\equiv |\psi_{2}|^2/|\psi_{1}|^2$. 
Now, $R$ ranges between one and two; when $R$ approaches either one on any of
the sites, we are in the ``localized regime'', while a value of two, indicates an
``extended regime''. From Eq.~(\ref{eq:6}) one obtains
\begin{equation}
\alpha = {\psi_{2}\over{\psi_{1}}} = {\lambda - \varepsilon_{1} - i \rho\over{C}}
\end{equation}
where $\lambda$ is given explicitly by Eq.~(\ref{eq:6b}). Figure~\ref{f2}(b)
shows $R$ vs $\rho$ for several ``disorder widths''. For a given disorder an
increase in gain and loss reduces $R$, while for a fixed gain and loss, an
increase in disorder also decreases $R$. It would seem that the presence of
gain and loss is effectively increasing the disorder, which reduces the spatial extent
of the stationary mode. 

\begin{figure}
\centering
\begin{overpic}[width=1\textwidth]{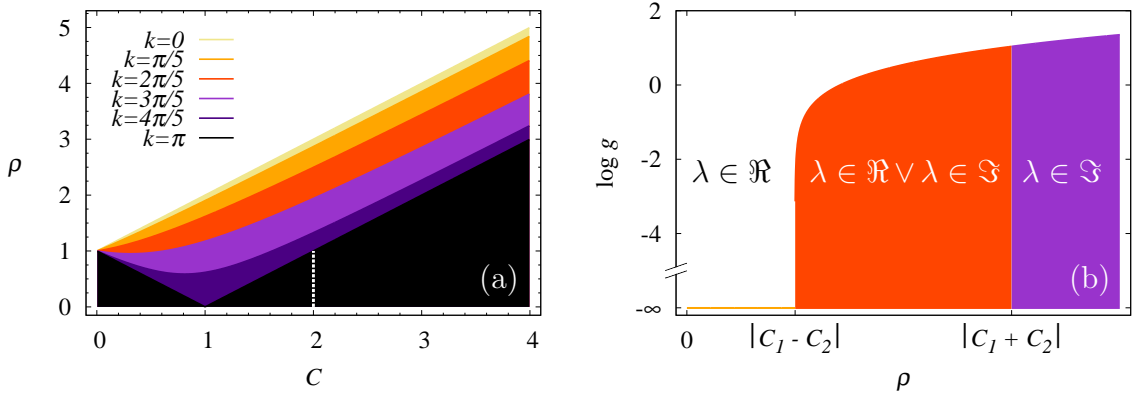}
\put(41,9){\white{(a)}}
\put(91,9){\white{(b)}}
\put(59,18){$\lambda\in\Re$}
\put(69,18){\white{$\lambda\in\Re$\,$\vee$\,$\lambda\in\Im$}}
\put(87,18){\white{$\lambda\in\Im$}}
\end{overpic}
\caption{(a) Stability regions as function of gain and loss parameter $\rho$
and coupling ratio $C$, for several wave vectors $k$.  Darkness increases with
$k$. Stable modes for all $k$ ($\lambda\in\Re$) can only exist within the
darkest region. (b) Instability gain $g$ (log scale) as a function of gain and
loss $\rho$, in the absence of disorder. The character of the eigenvalues changes with
$\rho$.}
\label{f3}
\end{figure}

Thus, from the results of the dimer model, we conclude that the interplay of
$\mathcal{PT}$-symmetry and disorder, tends to enhance the action of disorder,
while at the same time it leads the system into the broken
$\mathcal{PT}$-symmetry regime, for any amount of disorder.

\section{Long waveguide array}

\subsection{Gain and loss only}
Now we consider a long waveguide array with $N$ sites, with $N\gg 2$, described
by model~(\ref{eq1}). We consider first the case of absence of disorder ($W=0$),
but in the presence of the gain and loss. We look for stationary
modes of the form $\psi_{n,1(2)}(z) = \psi_{n,1(2)} e^{ik n + i \lambda z}$.
This leads to the linear equations
\begin{eqnarray}
(-\lambda + i \rho) \psi_{n,1} + (C_{1} e^{-i k}+ C_{2}) \psi_{n,2}&=&0,\nonumber\\
 (C_{2}+C_{1} e^{i k}) \psi_{n,1} + (-\lambda-i \rho) \psi_{n,2}&=&0
\end{eqnarray}
after imposing the condition that the determinant of the system be zero, in
order for nontrivial solutions to exist, we arrive at the dispersion relation
\begin{equation}
\lambda_{\pm}(k,\rho) = \pm \sqrt{\Delta} \label{eq12p},
\end{equation}
where $\Delta\equiv C_1^2+C_2^2-\rho^2+2C_1C_2\cos{k}.$
With this result we obtain the eigensolutions:
\begin{displaymath}
\left[ 
\begin{matrix} 
\psi_{1}^{\pm}, \\
\psi_{2}^{\pm} \\
\end{matrix}   
\right]=
\left[ 
\begin{matrix}
\delta_\pm, \\
1 \\
\end{matrix}
\right],\hspace{0.2cm}\mbox{where},\hspace{0.3cm}\delta_\pm=
\frac{i\rho\pm\sqrt{\Delta}}{C_2+C_1e^{ik}}.
\label{eq6}
\end{displaymath}
Stability domains or regions where the $\mathcal{PT}$-symmetry is preserved
correspond to values of $\lambda$ that are purely real. Inside the parameters
window where this occurs, there is balance between gain and losses in the
system. 

Fixing $C_1=1$, and defining $C\equiv C_{2}/C_{1}$, we can rewrite the
dispersion relation as
\begin{equation}			
\lambda_\pm(k,\rho)=\pm\sqrt{-\rho^2+1+C^2+2C\cos{k}}.
\label{eq12}
\end{equation}
In order to guarantee that $\lambda\in\Re$, the relation $\rho^2\leq
1+C^2+2C\cos{k}$ must be fulfilled for all wave number $k$.  Figure~\ref{f3}(a)
shows the stability regions in parameters space, the $\rnumber$ plane, for several
wave vectors $k$.  The different shaded areas represent stability domains for
several $k$ values. In particular there is a stability region valid for all $k$
values, shown as the darkest region in Fig.~\ref{f3}(a). This is the  most
important case, since when one considers the dynamical evolution of a general
optical excitation, each Fourier component will evolve according to one of the
eigenvalues; if one or several of some of them are  imaginary, the dynamics
will be unstable. Thus, for stability is necessary to stay inside the darkest
region in Fig.~\ref{f3}(a). It is also worth pointing out that for the case of a
homogeneous array, i.e., $C=1$, there is no absolute stability window for any
choice of parameters~\cite{Dmitriev:10}. Figure~\ref{f3}(b) shows the instability
gain defined  as the maximum of the absolute value of all the imaginary parts
of the eigenvalues. This instability gain will dominate the dynamics at long
propagation distances. Under the curve we have indicated the character of the
eigenvalues in different sectors of $\rho$ values.  For our normalization
choice, the first region with real eigenvalues only extends from $C=0$ up to
$C=1$. Between $C=1$ and $C=3$, the eigenvalues are either real or imaginary, 
and finally for $C>3$ the eigenvalues are all imaginary.  

Let us now go deeper into the localization of the light for systems
that exhibit a dispersion relation as from Eq.~(\ref{eq12p}). We start
by calculating the power content $P$ of the corresponding eigenmodes
\begin{equation} 
P_{\pm}=\sum_n^N{\left(1+\left|\delta_\pm\right|^2\right)}=\sum_{\rm{odd}}^N
1+\sum_{\rm{even}}^N
\left|\delta_\pm\right|^2=\frac{1+\left|\delta_\pm\right|^2}{2}N.
\label{eq7}
\end{equation}
Therefore, the participation ratio $R$ of an eigenmode is:
\begin{equation} 
R_{\pm}=\frac{\left(1+\left|\delta_\pm\right|^2\right)^2}{1+\left|\delta_\pm\right|^4}\frac{N}{2}.
\label{eq8}
\end{equation}
We have two cases to consider. The first one corresponds to $\Delta\geq 0$,
that is, inside the stable window. In that case, we have
\begin{equation} 
R_{\pm}=
\frac{\left[1+\left(\frac{i\rho\pm\sqrt{\Delta}}{C_2+C_1e^{ik}}\right)
\left(\frac{-i\rho\pm\sqrt{\Delta}}{C_2+C_1e^{-ik}}\right)\right]^2}
{1+\left[\left(\frac{i\rho\pm\sqrt{\Delta}}{C_2+C_1e^{ik}}\right)
\left(\frac{-i\rho\pm\sqrt{\Delta}}{C_2+C_1e^{-ik}}\right)\right]^2}\frac{N}{2}=N.
\label{eq9}
\end{equation}
In order to have an idea of the localization tendency of the whole system, we
proceed to take an average over all eigenmodes, that is, an average over all
wave vectors $k$:
\begin{equation}
\langle R_{\pm} \rangle^k=\frac{1}{2\pi}\int_0^{2\pi}R_{\pm}\rmd k=N.
\end{equation}
This means that the eigenmodes display complete delocalization in the
$\mathcal{PT}$-symmetry phase.  For the case $\Delta<0$, we are in broken
$\mathcal{PT}$-symmetry phase. The participation ratio is now

\begin{equation}
\eqalign{R_{\pm}&=
\frac{\left[1+\left(\frac{i\rho\pm i\sqrt{-\Delta}}{C_2+C_1e^{ik}}\right)
\left(\frac{-i\rho\mp i\sqrt{-\Delta}}{C_2+C_1e^{-ik}}\right)\right]^2}
{1+\left[\left(\frac{i\rho\pm i\sqrt{-\Delta}}{C_2+C_1e^{ik}}\right)
\left(\frac{-i\rho\mp i\sqrt{-\Delta}}{C_2+C_1e^{-ik}}\right)\right]^4}\frac{N}{2},\cr
&=\frac{-N\rho^2}{C_1^2+C_2^2-2\rho^2+2C_1C_2\cos{k}},}
\label{eq10}
\end{equation}
and the mean participation ratio will be given by
\begin{equation}
\langle R_{\pm} \rangle^k=\frac{N}{2\pi}
\int_0^{2\pi}\rmd k\frac{-\rho^2}{C_1^2+C_2^2-2\rho^2+2C_1C_2\cos{k}}.
\label{eq11}
\end{equation}

Equation~({\ref{eq11}}) establishes, in a closed form, the evolution of the
participation rate for a binary lattice in terms of the strength of gain and loss
parameter, as well as a function of the strength of its couplings.  
Figure~\ref{f4}(b) (uppper curve) shows  $\langle R \rangle^k$ as a function of
$\rho$, in the absence of disorder. As $\rho$ increases, $\langle R \rangle^k$
decreases, indicating a greater localization. This is reminiscent of Anderson
localization with $\rho$ playing the part of the disorder width, which is in
qualitative agreement with recent experiments~\cite{ptncom}.

\subsection{Gain and loss plus disorder}

\begin{figure}
\centering
\begin{overpic}[width=1\columnwidth]{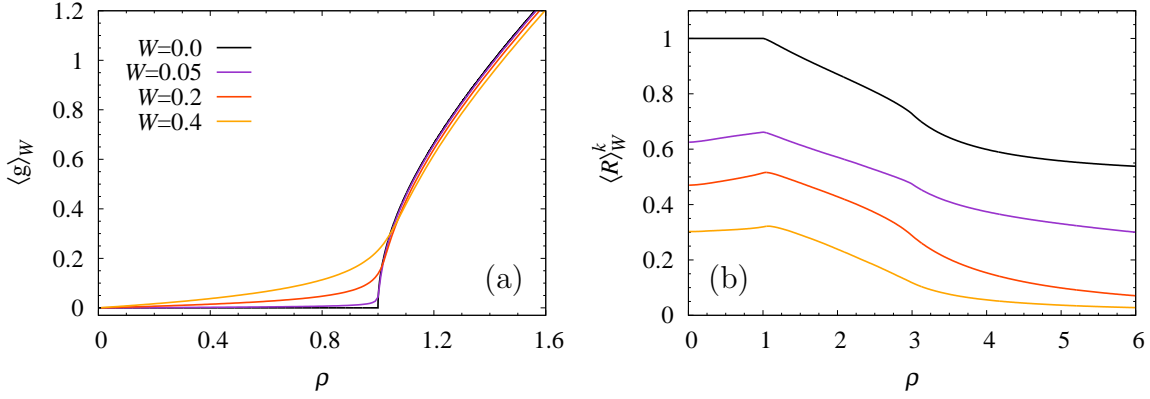}
\put(41,9){(a)}\put(60,9){(b)}
\end{overpic}
\caption{ (a) Averaged instability gain $\langle g\rangle_W$ and (b)
participation ratio $R$ in a 1D binary array with disorder, gain and loss, as a
function of the gain and loss parameter $\rho$, both for several disorder
widths labeled at the inset. $C=2$.}
\label{f4}
\end{figure}

Let us now add disorder into the picture. The presence of disorder makes the
system no longer $\mathcal{PT}$-symmetric, and some eigenvalues will be
complex. Disorder also destroys the periodicity of the system and the
computation of its eigenvalues and eigensolutions must proceed numerically.
The instability gain, $g\equiv\Im{(\lambda)}_{\rm{max}}$, will dominate the
dynamics at long propagation distances. Figure~\ref{f4}(a) shows this
instability gain as a function of the gain and loss parameter, for several
disorder widths labeled at the inset, and a coupling ratio of $C=2$.  In general, for a given
disorder width, the gain increases monotonically with $\rho$, converging
eventually to the curve $g(\rho)=\sqrt{\rho^2-1}$. On the other hand, for a
fixed $\rho$, the gain also increases with disorder. This behavior of the gain
suggest that the presence of both, disorder and gain and loss, tend to
destabilize the system. In Fig.~\ref{f4}(b) we show the participation ratio,
this time averaged over all eigenstates and over a number of disorder
realizations ($N_r=100$). This double-averaged parameter serves as an estimator
for the localization tendency of the system. As we can see, for a fixed gain
and loss value, an increase in disorder decreases $\langle R \rangle^k_W$,
indicating an increase in localization, as expected on general grounds. On the
other hand, for a fixed disorder, $\langle R \rangle^k_W$ first increases with
$\rho$, reaches a maximum, and finally decreases steadily with further increase
in $\rho$. Note that the maximum occurs at $|C_{1}-C_{2}|=1$, and that there is
an inflection point at $|C_{1}+C_{2}|=3$. We have seen these two special points
before when examining the instability gain in the absence of disorder
[Fig.~\ref{f3}(b)]. Now, the initial increase of $\langle R \rangle^k_W$ with
$\rho$ indicates that, as $\rho$ is increased, the optical power content of the
modes becomes more uniformly distributed in space. A very similar phenomenon
has been observed in lattices with disorder and nonlinearity~\cite{Naether:13}.

\section{Transport properties}

Let us now consider the problem of the transport of optical power in this
binary waveguide array modeled by Eq.~($\ref{eq1}$), originally
$\mathcal{PT}$-symmetric, and then slightly perturbed by introducing disorder
into their propagation constants, that is by imposing a random distribution of
indices of refraction. We will focus on two cases: the propagation of initially
localized (delta function-like) and of extended (plane wave) excitations.  

\subsection{$\mathcal{PT}$ vs disorder for delta-like beam excitation}

We start by analyzing the dynamical evolution of a narrow input beam focused on
the central guide of the array.  For that we integrate numerically the
model~(\ref{eq1}), for a binary waveguide array in the presence of alternating
gain and losses and linear disorder. We will focus on the mean size of the
wave-packet upon the beam propagation, measured by the mean square
displacement,
\begin{equation} 
M_2={\sum_{i=1}^{N}(n-n_c)^2\left|\psi_i\right|^2\over{\sum_{i=1}^{N} |\psi_{i}|^2}}. 
\label{eq13}
\end{equation}
In our simulations, we take $N=1200$, and $n_c=N/2$ is the initially excited
waveguide. It is worth mentioning that model~(\ref{eq1}) (for $\rho\neq
0$) is a non-Hermitian system, then, there is no conserved quantities (integrals
of motion) during propagation. For instance, the optical power $P=\sum_{n}
|\psi_{n}|^2 $ is not a dynamical constant and we expect that, in the absence
of disorder, its value will oscillate.  However, in the presence of disorder
the $\mathcal{PT}$-symmetry could be broken leading to the growth of the
optical power.

Since we are dealing with disordered arrays, we must  collect information from
a number $N_r$ of different disorder realizations, and then take the average
over them.  Quantities (\ref{eq2}) and (\ref{eq3}) are also useful in that they
tell us how the light is distributing along the array upon propagation. In the
following numerical analysis, we have set a coupling ratio of $C=2$ and, for
each case, we perform one hundred disorder realizations ($N_r$=100). In the
absence of disorder, the $\mathcal{PT}$-symmetry will hold for $\rho\leq1$ [see
white dotted line in Fig.~\ref{f3}(a)]. However, the interplay between gain and
losses with disorder breaks the $\mathcal{PT}$-symmetry, could lead to the
emergence of eigenfunctions with complex eigenvalues.

\begin{figure}
\centering
\begin{overpic}[width=1\textwidth]{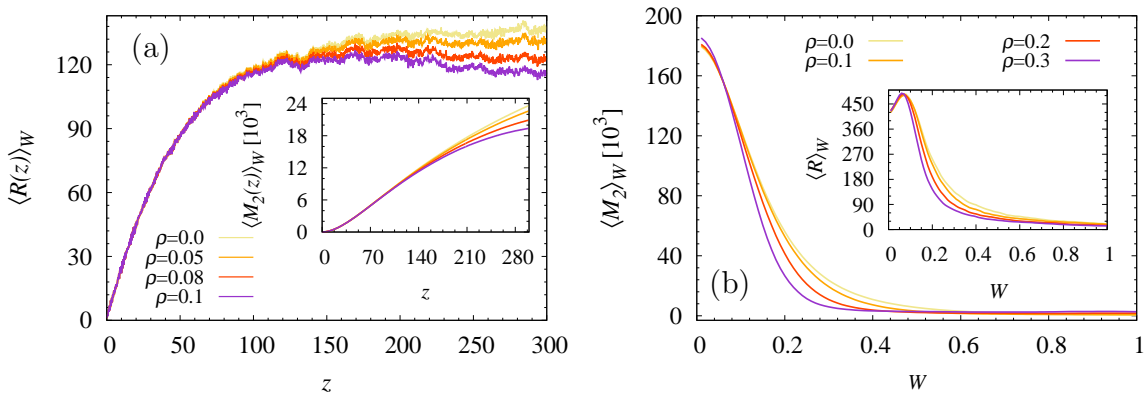}
\put(11,29){(a)}\put(60,9){(b)}
\end{overpic}
\caption{Dynamical wavepacket evolution in 1D binary array with disorder,
gain and loss. (a) $\langle R(z) \rangle_W$ and $\langle
M_2(z) \rangle_W$ (inset) are displayed for several gain and loss values in the
range $0\leq\rho\leq 0.1$, and (b) $\langle M_2 \rangle_W$ and $\langle R \rangle_W$
(inset) as a function of $W$ for several gain and loss values in the
range $0\leq\rho\leq 0.3$. $C=2$.}
\label{f5}
\end{figure}

Figure~\ref{f5}(a) displays four cases of $\langle R(z) \rangle_W$ evolution for
disordered binary arrays of length $z=300$. Each of them corresponds to a
different value of $\rho$ parameter but keeping the same width of disorder
$W=0.3$. The brightest line stands for $\rho=0$, i.e., in the absence of gain and
losses. The other lines correspond to $\rho=0.05, 0.08$ and
$0.1$, respectively. From here, we clearly see how $\langle R(z) \rangle_W$ tends
to saturate due to wavepacket localization, in agreement with the thesis of
Anderson. Nevertheless, the number of effective excited sites diminish with
the increment of $\rho$ values, i.e., the presence of gain and losses
alternately distributed contributes to localize the wavepacket further.
Similarly, from the inset in Fig.~\ref{f5}(a) we observe that $\langle M_2(z)
\rangle_W$ also evolves towards a saturation as expected from Anderson
localization. 

Figure~\ref{f5}(b) shows the effect of disorder on the width of the wavepacket
$\langle M_{2}(z)\rangle_W$ at the output of an array of length $z=100$, for
several values of the gain and loss parameter. In all cases, as the width of
the disorder increases, $\langle M_{2}(z)\rangle_W$ decreases steadily, as a
power law. This decrease is faster for larger values of $\rho$. The behavior of
the average participation ratio as a function of disorder, displayed as an
inset in Fig~\ref{f5}(b), show the same behavior, except at small disorder
widths where $R$ increases with $W$, for all $\rho$. We have noticed a similar
behavior for $R$ when we discussed Fig.~\ref{f3}. In other words, for small
disorder widths there is a tendency to redistribute the optical power content
in a more uniform manner among the guides~\cite{Naether:13}.

\subsection{$\mathcal{PT}$ vs disorder for an extended beam excitation}

\begin{figure}
\centering
\begin{overpic}[width=0.85\textwidth]{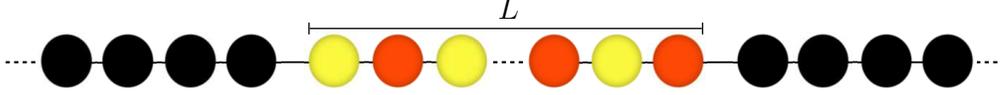}
\psline[linewidth=0.5pt]{|-|}(3.75,0.85)(9.,0.85)\put(47,7.3){$L$}
\end{overpic}
\caption{Sketch of a disordered segment, of length $L$, with alternating gain
(yellow filled circles) and losses (orange filled circles), embedded in a 1D
linear homogeneous lattice (black filled circles).} 
\label{f6}
\end{figure}

Finally, we analyze the averaged transmission $\langle T \rangle_W^k$ of a
plane wave across a disordered segment  of length $L$ containing gain and
losses, as well as disorder. We assume the segment embedded in a large
homogeneous 1D lattice (black filled circles).  An sketch of the system is
shown in Fig.~\ref{f6}, where orange (yellow) filled circles represent
those sites with losses (gain).

We are interesting in knowing how the transmissivity, as a function of $L$, is
affected by the interplay of disorder and the presence of gain and loss. In the
absence of gain and loss, it is well-known that the transmission would decay
exponentially with the size of the disordered segment~\cite{economou2006green}.
When disorder and nonlinearity are present, the transmission decays as a
power-law~\cite{PhysRevB.58.12547}.

Outside the ``impurity'' segment, the system is  modeled by the discrete
Schr\"odinger equation,
\begin{equation}
\frac{d\psi_n}{dz}+V(\psi_{n-1}+\psi_{n+1})=0,
\label{eq14}
\end{equation}
which has stationary solutions of the form $\psi_{n}=\phi_{n}e^{i\lambda z}$,
leading to the dispersion relation $\lambda=2V\cos{k}$.  On the other hand,
inside the segment, the field is governed by model (\ref{eq1}), which can be
re-written in the following way:
\begin{equation}
i \frac{ d\psi_n}{dz}+C_{n,n-1}\psi_{n-1}+C_{n,n+1}\psi_{n+1}+\gamma_n\psi_n=0,
\label{eq15}
\end{equation}
where now $\gamma_n=1+\varepsilon_n\pm i\rho_n$, with $\rho_{n}=\rho$ $(-\rho)$ for $n$
even (odd). Its stationary version is given by
\begin{equation}
\lambda\phi_n+C_{n,n-1}\psi_{n-1}+C_{n,n+1}\psi_{n+1}+\gamma_n\psi_n=0.
\label{eq17}
\end{equation}
Let us now consider the transmission of an extended excitation, i.e., a plane
wave across the segment:
\begin{eqnarray}
\psi_n = \left\{
\begin{array}{l@{\quad}l}
R_0e^{ikn}+R_1e^{-ikn},\, n \leq 0,\\
R_2e^{ikn},\, n \ge L.
\end{array}
\right.
\label{eq16}
\end{eqnarray}
From Eq.~(\ref{eq17}), we obtain the recurrence relation
\begin{equation}
\psi_{n-1}=\frac{(\lambda-\gamma_n)\psi_n-C_{n,n+1}\psi_{n+1}}{C_{n,n-1}},
\label{eq18}
\end{equation}
which we will use to compute the transmission: for a given wave vector $k$, one
starts at the end of the segment $n=L$ and assumes a given value for $R_{2}$.
For example $R_{2}=1$. Therefore, from Eq.~(\ref{eq16}), at $N=L$ and $n=L+1$,
$\psi_{L}=\exp(ikL)$ and $\psi_{L+1}=\exp[i k (L+1)]$, respectively. Then we
iterate backwards using the above recurrence relation, Eq.~(\ref{eq18}), until
we reach the beginning of the segment where $R_0$ is computed. The
transmissivity is then given by $T=\left|R_2\right|^2/\left|R_0\right|^2$.

\begin{figure}
\centering
\begin{overpic}[width=1\columnwidth]{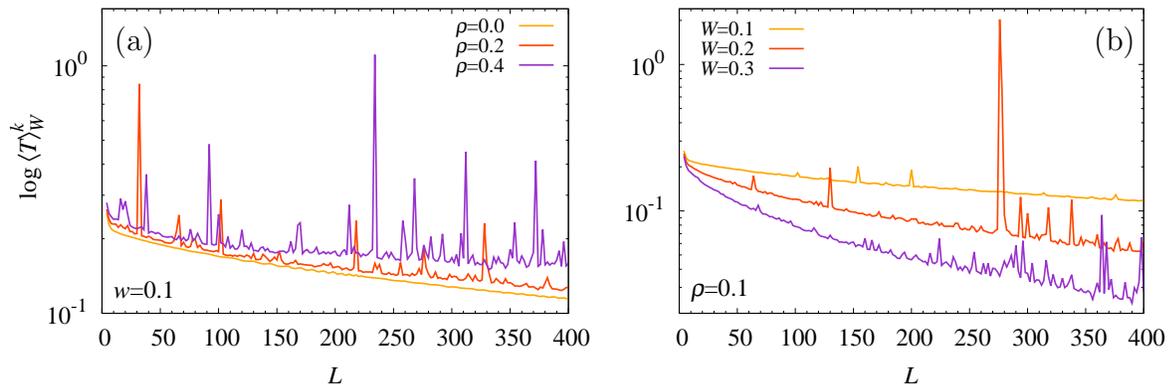}
\put(9,29){(a)}
\put(92,29){(b)}
\end{overpic} 
\caption{Averaged transmission $\langle T \rangle_W^k$
of a plane wave across a disordered segment containing  gain and loss, as
function of the length of the segment $L$. (a) $\rho=0.0$ ($\rho=0.2$ and
$\rho=0.4$) yellow (orange and purple) line for fixed $W=0.1$, and (b) $W=0.1$
($W=0.2$ and $W=0.3$) yellow (orange and purple) line for fixed
$\rho=0.1$.}
\label{f7}
\end{figure}

Figure~\ref{f7} shows the average transmission (log scale) across a disordered
segment of length $L$, with gain and losses. We have averaged over one hundred
disorder realizations, and also over all wavevectors $k$. In general, we see
that $\langle T \rangle_W^k$ decreases with $L$, and this tendency is stronger
when the width of disorder increases. This is shown in Fig.~\ref{f7}(b), where
light-gray (gray and dark-gray) lines correspond with $W=0.1$ ($W=0.2$ and
$0.3$).

Figure~\ref{f7}(a) shows something interesting: as the gain and loss
coefficient is increased (for fixed disorder and fixed $L$), the transmission
increases with $\rho$. This is in marked contrast to the case of the delta-like
beam excitation where the opposite tendency occurred.  We also observe the
presence of fluctuations which can be quite strong in some cases like in
$W=0.2$, $\rho=0.1$, and also in $W=0.1$, $\rho=0.4$. They seem to be generic
and do not change significantly with finer  wavevector discretization. 

\section{Discussion}

We have examined the transport  of excitations across a 1D binary lattice, in
the presence of disorder, plus the presence of gain and loss. In the absence of
disorder, the system is $\mathcal{PT}$-symmetric. As a first approach to the
problem we studied a dimer reduction, observing that the interplay of
$\mathcal{PT}$-symmetry and disorder, tends to enhance the action of disorder,
while at the same time it leads the system into the broken
$\mathcal{PT}$-symmetry regime, for any amount of disorder. Next, we examine
the case of a long binary lattice, finding that as soon as disorder is introduced, the
system goes into the broken $\mathcal{PT}$-symmetry phase, and that the
presence of gain and loss tends to reinforce the action of disorder.  
 
Next we consider the propagation of localized and extended excitations inside
the binary system. For the case of the delta-like initial beam, we observe that
its propagation is somewhat inhibited by an increase in gain and loss.
Surprisingly, the opposite happens when examining the transmission of plane
waves across  a binary lattice segment with disorder and gain and loss: in that
case, the presence of gain and loss tends to increase the transmission.  This
transmission experiments robust fluctuations overimposed over its well-defined
decaying behavior as the segment length increases.  These fluctuations appear
independent of the width of disorder or the strength of gain and loss
parameter. Moreover, we have observed fluctuations for the case of a fixed
$\rho$ and $L$ and varying disorder $W$.  We believe that the origin of these
fluctuations with $L$ or $W$ have their origin in the complex eigenvalue
spectra of the system. For a fixed $\rho$ and $L$, the set of eigenvalues will
change from random realization to realization, introducing new instability
gains which might cause the transmission to change abruptly. On the other hand,
for a system with fixed disorder and gain and loss, a change in $L$, generates
a different set of complex eigenvalues where, again, the instability gain might
change, even for as small a change as one site. The fluctuations can become so
strong as to generate transmissions greater than unity (see Fig~\ref{f7}).
 
We conclude that, for a binary chain,  the interplay of disorder and gain and
loss tends to reduce the spatial extent of the eigenmodes and that it favors
(inhibits) the dynamical propagation of extended (localized) excitations,
giving also rise to strong fluctuations in the transmission of plane waves
across the system. 

\section{Acknowledgments}
 
This work was supported in part by Fondo Nacional de Ciencia y Tecnolog\'{\i}a
(Grants 3140608 and 1120123), Programa Iniciativa Cient\'{\i}fica Milenio
(Grant P10-030-F), Programa de Financiamiento Basal (Grants FB0824 and
FB0807) and by the supercomputing infrastructure of the NLHPC (ECM-02).

\section*{References}


\providecommand{\newblock}{}

\end{document}